\documentclass[twocolumn,showpacs,preprintnumbers,amsmath,amssymb,prl,aps]{revtex4}

\usepackage{graphicx}
\usepackage{dcolumn}
\usepackage{bm}

\begin{document}

\title{Origin of charge density wave formation in insulators
from a high resolution photoemission study of BaIrO$_3$}

\author{Kalobaran Maiti, Ravi Shankar Singh, V.R.R. Medicherla, S. Rayaprol, and
E.V. Sampathkumaran}

\affiliation{Department of Condensed Matter Physics and Materials'
Science, Tata Institute of Fundamental Research, Homi Bhabha Road,
Colaba, Mumbai - 400 005, INDIA}

\date{\today}

\begin{abstract}

We investigate the origin of charge density wave (CDW) formation
in insulators by studying BaIrO$_3$ using high resolution (1.4
meV) photoemission spectroscopy. The spectra reveal the existence
of localized density of states at the Fermi level, $E_F$, in the
vicinity of room temperature. These localized states are found to
vanish as the temperature is lowered thereby, opening a {\em soft
gap} at $E_F$, as a consequence of CDW transition. In addition,
the energy dependence of the spectral density of states reveals
the importance of magnetic interactions, rather than well-known
Coulomb repulsion effect, in determining the electronic structure
thereby implying a close relationship between ferromagnetism and
CDW observed in this compound. Also, Ba core level spectra
surprisingly exhibit an unusual behavior prior to CDW transition.
\end{abstract}


\maketitle

The investigation of charge density wave (CDW) and its influence
on the properties of materials has been the subject of intense
research for many decades now. Such a phenomenon is often observed
in low-dimensional metallic systems due to the spatial modulation
of the conduction electron density, the periodicity of which is
not commensurate with the unit-cell. As a consequence, a gap opens
up in the single particle excitation spectrum of such materials
below the transition temperature. Clearly, CDW  is expected only
for metallic systems. Interestingly, BaIrO$_3$, an insulating
material, exhibits CDW transition at about 175~K \cite{cao1}. As
such an unusual observation is likely to evoke considerable
interest, it is important to understand the origin of the CDW
anomaly in such insulating materials.

BaIrO$_3$ exhibits many interesting crystallographic and magnetic
anomalies \cite{cao1,cao2,powell,lindsay,chamberland}. The crystal
structure is monoclinic (space group C2/m) and consists of
Ir$_3$O$_{12}$ trimers, where the IrO$_6$ octahedra are face
shared. Inter-trimer link occurs by corner-sharing to form columns
parallel to $c$-axis, as shown in the inset of Fig.~1. It appears
\cite{lindsay} that the trimers are marginally tilted with respect
to each other and thus a moment-containing chain runs parallel to
$c$-axis. Quasi-one-dimensional nature of this compound is evident
from the anisotropic transport behavior \cite{cao1}. The
complexity of the structure results in twisting and buckling of
the trimers, and in multiplicity of Ir-O and Ba-O bond distances,
thereby creating 4 types of Ir and 3 types of Ba
\cite{powell,siegrist}. A finding unique to this compound is that
ferromagnetism sets in at the same temperature at which CDW forms!
Thus, it is fascinating that there is an intimate relationship
between magnetism and lattice. Therefore, there is an urgent need
to throw more light on this relationship. Many other interesting
properties are exhibited by this compound, for instance, very low
magnetic moment, additional transitions around 80 and 26~K,
non-metallicity despite very short Ir-Ir distances etc. All these
anomalies reveal that BaIrO$_3$ is truly an exotic material,
warranting further urgent investigations.

In this letter, we provide experimental evidence for the fact that
there are indeed electronic states at the Fermi level, $E_F$,
which are localized presumably due to crystallographic features
like disorder and distortions, and that the CDW evolves due to
these localized electronic states. While the existence of such
localized states forming a 'pseudogap' at $E_F$ has been predicted
long ago \cite{mott,anderson}, it is the high resolution employed
in the present study that enables us to show directly the presence
of such states and to track its temperature dependence to
understand the CDW in insulators. $|E-E_F|^{3/2}$-dependence of
the spectral density of states (SDOS) emphasizes the role of
magnetism in determining the electronic structure. In addition, we
observe significant changes in Ba core level spectra prior to CDW
transition suggesting an importance of Ba-O covalency in its
electronic properties.

The sample was prepared by a conventional solid state reaction
method in the polycrystalline form using ultra-high pure
ingredients (BaCO$_3$ and Ir metal powders). In order to achieve
large grain size and stoichiometry, the final product was
pelletized and sintered at a rather high temperature (1000~$^o$C)
for more than 2 days and furnace-cooled (rather than quenching)
which ensures the oxygen stoichiometry close to 3
\cite{chamberland}. The $x$-ray diffraction pattern exhibits
single phase without any signature of impurity and the lattice
constants ($a$ = 10.015~\AA, $b$ = 5.748~\AA, $c$ = 15.157~\AA\
and $\beta$= 103.27$^o$) are found to be in excellent agreement
with those reported for single crystals \cite{cao1}. In addition,
we performed electrical resistivity measurements by a conventional
four-probe method and establish the presence of a CDW transition
at 183~K from an insulating phase; the onset of a ferromagnetic
transition at the same temperature is observed by dc magnetic
susceptibility measurements taken in the presence of a magnetic
field of 5~kOe. Somewhat higher transition temperature observed in
this compound compared to the single crystals \cite{cao1} ensures
good quality of the sample. Photoemission measurements were
performed using a Gammadata Scienta analyzer, SES2002 at a base
pressure of 4$\times$10$^{-11}$~torr. The experimental resolution
was 0.8~eV for Al~K$\alpha$ (1486.6~eV), 4.5~meV for
He~{\scriptsize II} (40.8~eV) and 1.4~meV for He~{\scriptsize I}
(21.2~eV) measurements. The sample surface was cleaned by {\em in
situ} scraping and the cleanliness was ascertained by tracking the
sharpness of O~1$s$ feature and absence of C~1$s$ peaks.

Valence band spectra at room temperature obtained using 21.2~eV,
40.8~eV and 1486.6~eV excitation energies are shown in Fig. 1.
There are four discernible features marked by A, B, C and D in the
figure. While features B and C appearing at binding energies (BE)
6~$>~BE~>$~3 eV are large in the low excitation energy spectra,
the features A and D are enhanced significantly in the
Al~$K_\alpha$ spectrum. The intensity of the feature D becomes
maximum in the latter. Considering strong dependence of the
relative transition matrix elements on excitation energies
\cite{yeh}, the features B and C can be attributed to O~2$p$
non-bonding electron excitations and the feature A to the bonding
levels with the electronic states having primarily O~2$p$
character. The feature D corresponds to the excitation of
primarily Ir 5$d$ electrons.

Since a major conclusion is based on the spectra obtained at low
excitation energies, it is important to establish that our
conclusions are representative of the bulk. The feature D in
Fig.~1 is distinctly separated from the O~2$p$-related signals in
the 40.8~eV and 1486.6~eV spectra. Thus, Ir~5$d$ contributions can
reliably be delineated by fitting the O~2$p$ bands using three
Gaussians representing features A, B and C as performed in other
systems \cite{casrvo,lacavo,kbmprb,ruthenates}. The resultant fit
obtained by least square error method is shown by lines in Fig.~1
and the extracted Ir~5$d$ band is shown in Fig.~2(a). The spectral
intensity at the Fermi level appears to be significantly small in
both the spectra, and a peak appears around 1.3~eV binding energy
with a tail extending down to 3~eV. Interestingly, the spectral
lineshape of both 40.8~eV and 1486.6~eV-spectra are identical
despite their large difference in probing depth; this is
demonstrated by superposing the resolution broadened 40.8~eV
spectrum (solid line in the figure) over the 1486.6~eV-spectrum.
This establishes that the surface and the bulk electronic
structures are essentially identical in contrast to the
observations in 3$d$ and 4$d$ transition metal oxides
\cite{casrvo,lacavo,kbmprb,ruthenates}.

We now focus on the evolution of the valence band spectral
intensities as a function of temperature in order to investigate
the origin of CDW ground state to bring out the point of major
emphasis of this paper. No spectral modification is observed down
to 183~K. As the temperature is lowered across 183~K, the
intensity at E$_F$ in the 40.8~eV spectra decreases dramatically
and the leading edge shifts to higher binding energies as shown in
Fig.~2(b). Large resolution broadening in the 1486.6~eV spectra
smears out all these changes; however, these spectra ensure that
there is no significant modification at higher energy scales. In
order to bring out a better clarity, we expand $E_F$-region of
40.8~eV spectra in Fig.~2(c) at temperatures 300K, 150K and 28K.
The shift of the leading valence band edge to higher binding
energy is clearly evident at 150~K, which reduces the spectral
weight at $E_F$ significantly. A further decrease in temperature
leads to an opening of a band gap of the order of 50~meV below
$E_F$ accompanied by an enhancement in intensity around 0.9 eV
binding energy as inferred from Fig.~2(b).

In order to discuss the changes at $E_F$ in more detail, we show
the high resolution spectra obtained with He~{\scriptsize I}
radiations in Fig.~3. The shift in the valence band edge with
decreasing temperature is clearly evident in Fig.~3(a). The
photoemission response can be expressed as $I(\epsilon) = \int
g(\epsilon_1).F(\epsilon_1,T).
L^e(\epsilon_1,\epsilon_2).L^h(\epsilon_2,\epsilon_3).
G(\epsilon_3,\epsilon)d\epsilon_1 d\epsilon_2 d\epsilon_3$ where,
$g(\epsilon)$ and $F(\epsilon_1,T)$ represent the matrix element
weighted density of states and Fermi-distribution function,
respectively. The Lorentzian broadenings ($L^e$ and $L^h$) due to
the finite lifetime of photo-electrons and photo-holes will be
small close to $E_F$. Since the resolution broadening represented
by the Gaussian, $G$ (full width at half maximum = 1.4 meV) is
negligible compared to the energy scale of investigation,
$I(\epsilon)/F(\epsilon,T)$ provide a good representation of the
spectral density of states, SDOS, as has also been observed in
other systems \cite{kbmprb}.

The SDOS thus obtained at 300~K and 150~K are shown in Fig.~3(b)
along with the spectra at other temperatures. It is important to
note here that such an estimation of SDOS is sensitive to the
precise location of $E_F$. Therefore, we have carefully determined
$E_F$ at each temperature by the Fermi cut off as observed for
silver mounted on the sample holder together with the sample. The
representative spectra are shown in Fig.~3(a) at 20~K and 300~K.
In order to verify the reliability of our analysis, we carry out
the same exercise for silver as described above (see Fig.~3(b)).
The room temperature spectrum clearly reproduces the flat density
of states of silver in a wide energy range ($>$~3.5~$k_BT$),
thereby providing a confidence in this procedure.

Interestingly, it is clear from Fig.~3(b) that the spectral DOS at
room temperature exhibits a finite intensity and a distinct dip at
$E_F$ suggesting the signature of a {\em 'pseudogap'} \cite{mott}.
This observation in conjunction with the insulating transport
behavior \cite{cao1} suggests that all these electronic states are
essentially localized. This is not surprising for such a quasi-one
dimensional system where a small lattice distortion and/or
impurity leads to localization of the electronic states. Existence
of such localized states forming a {\em pseudogap} at $E_F$ has
been predicted long before \cite{mott,anderson}. High resolution
employed for these measurements makes it possible to directly
probe such electronic states experimentally. A notable finding is
that the intensity at $E_F$ becomes close to zero below the CDW
transition, thus forming a {\em soft gap} at $E_F$. Eventually a
gap of the order of 50~meV below $E_F$ opens up around 83~K with a
corresponding increase in intensity around 0.9~eV binding energy.

It is well known that the disorder leads to a $|E-E_F|^{1/2}$ ($E$
is energy) cusp at $E_F$ in metals \cite{altshuler-aronov,ddsprl}.
In an insulator consisting of localized electronic states at
$E_F$, a soft Coulomb gap opens up due to electron-electron
Coulomb repulsion; in such a situation, the ground state is stable
with respect to a single-particle excitation only if SDOS can be
characterized by $(E-E_F)^2$-dependence \cite{efros,massey}. In
contrast to this, in the present case, the SDOS exhibits a
$|E-E_F|^{3/2}$-dependence spanning a large energy range close to
$E_F$ (BE~$\leq$~300~meV) as shown in Fig.~3(c). Most
interestingly, this dependence remain unchanged down to the lowest
temperature studied. Thus, the opening of a {\em soft gap}
observed here has an origin different from the electron-electron
Coulomb repulsion observed in other systems \cite{massey}. The
exponent of 3/2 in SDOS suggests strong influence of
electron-magnon coupling on the electronic structure \cite{irkhin}
presumably responsible for the ferromanetic ground state.

We now turn to another fascinating observation in the core level
spectra of metallic ions. In Fig.~4(a), we show Ba 3$d_{5/2}$
spectra exhibiting a sharp feature around 779~eV binding energy.
Interestingly, a decrease in temperature from 300~K to 210~K leads
to a significantly large shift in the peak position and {\it there
is no further shift} as the temperature is lowered further
\cite{sampath2}. Similar effect has also been observed for other
core levels as shown for Ba 4$d_{5/2}$ in Fig.~4(b), while the Ir
core levels remain unchanged (see the same figure). The linewidth
of the curves for Ba remain the same below 210~K but it is much
larger at room temperature. Interestingly, 300~K spectra can be
reproduced remarkably by a superposition of the 25~K spectra
($\sim$~78\%) and the same shifted by 0.5~eV ($\sim$~22\%) towards
higher binding energy as shown in the figure. It is therefore
clear that the spectra for Ba at room temperature consists of
signals from two types of Ba ions. However, there are three types
of Ba ions in the crystal structure \cite{powell} depending upon
the Ba-O bond distances and clustering. We, thus, calculated the
Madelung potential considering only the nearest neighbor cluster
($V = \sum (a/r_{i})$; $a$ is a constant). We find that this
potential is nearly the same ($\sim$4.05$a$) for two types of Ba
ions, which is slightly different from the third one
($\sim$4.1$a$), thereby, qualitatively accounting for the
features. In short, the above results reveal that there are
profound changes in the covalency effects around Ba at a
temperature higher than the CDW transition temperature. To our
knowledge, it is for the first time that such an effect is
identified prior to the formation of CDW. While it is not clear
whether this is a precursor effect to CDW formation, a
confirmation of such a relationship from future studies is
expected to open up new direction to understand the origin of CDW,
just as pseudo-gap effects in high temperature superconductors
have attracted a lot of attention. We hope this work will motivate
further work to look for similar features in other non-metals,
which will help in the advancement of the knowledge of phase
transitions.

In conclusion, to our knowledge for the first time, we have
demonstrated directly the evolution of charge density wave due to
localized electronic states. It is also shown that a {\em soft
gap} opens up across the CDW transition. The energy dependence of
the spectral density of states reveals the role of magnetism on
the electronic structure in the vicinity of Fermi level, which
signals an intimate relationship between ferromagnetism and charge
density wave in this system. The observation of profound changes
in the Ba-O covalency prior to the formation of charge density
wave poses a new question with respect to the role of precursor
effects.

\section{Figure Captions}

Fig. 1 Valence band spectra of BaIrO$_3$ at different excitation
energies. The line represents the simulated O~2$p$ contributions
using three Gaussians representing features A, B and C. The inset
shows two Ir$_3$O$_{12}$ clusters, which are connected by corner
sharing with the cluster long axis tilted by 12$^o$ leading to
monoclinic structure.

 Fig. 2(a) Ir~5$d$ contributions in Al~$K\alpha$ (open circles) and
He~{\scriptsize II} (solid circles) spectra after subtraction of
the O~2$p$ features shown in Fig.~1. The solid line through
Al~$K\alpha$ spectrum, represents the resolution broadened
He~{\scriptsize II} spectrum. (b) The temperature evolution of the
valence band at Al~$K\alpha$ and He~{\scriptsize II} excitation
energies. He~{\scriptsize II} spectra show significant shift of
the leading edge at $E_F$ with decreasing temperature and the
spectral weight appears to be transferred from $E_F$ to around
0.9~eV binding energy. (c) Expanded He~{\scriptsize II} spectra
close to $E_F$ clearly exhibiting the shift of the leading edge
across CDW transition. The 25~K spectrum shows a gap of about
50~meV below $E_F$.

Fig. 3(a) High-resolution He~{\scriptsize I} spectra near $E_F$.
Ag spectra at 20~K and 300~K are also shown by stars to precisely
determine $E_F$. (b) Spectral density of states (SDOS) at room
temperature (open circles) and at 150~K (up-triangles) are
compared with the spectra at other temperatures. SDOS for Ag at
300~K is shown by stars. (c) SDOS plotted as a function of
$(BE)^{1.5}$ show linear dependence.

Fig. 4(a) Ba 3$d_{5/2}$ and (b) Ba 4$d_{5/2}$ core level spectra
at different temperatures. The dashed lines A(C) and B(D) in
upper(lower) panel are the spectra at 25~K and the same shifted by
0.5~eV (0.45~eV) towards higher binding energy, respectively. The
weighted sum, A+B (C+D) reproduce the 300~K spectrum (solid line)
remarkably well. Ir 4$f$ spectra collected in the same scan with
4$d$ spectra are shown in (b), which do not show any change with
temperature.


\begin{thebibliography}{}
\bibitem{sampath2} A careful look at the first and second derivative of
electrical resistivity vs temperature revealed the existence of an
anomaly around 250 K and it is of interest to confirm this on a
variety of samples.
\end{thebibliography}
\end{document}